\documentstyle [epsf]{article}

\makeatletter
\@addtoreset{equation}{section}
\makeatother

\begin{document}

\title
{The Degree of Generality of Inflation in FRW Models with
 Massive Scalar Field and Hydrodynamical Matter}

\author{A.V. Toporensky}

\date{}
\maketitle
{Sternberg Astronomical Institute, Moscow University, Moscow 119899, Russia}

\abstract
{Friedmann-Robertson-Walker cosmological models with a massive scalar
field are studied in the presence of hydrodynamical matter
in the form of a perfect fluid. The ratio of the number of solutions
without inflation to the total number of solutions is evaluated, depending on
the fluid density. It is shown that in a closed model this ratio can reach
$60\,\%$, by contrast to $\sim 30\,\%$ in models without fluid. \\
PACS  {05.45.+b 98.80.Hw 98.80.Cq} }

\section{Introduction}

In the recent two decades the dynamics of an isotropic Universe filled with
a massive scalar field attracted a great attention \cite{Linde}. From
the physical point of view, the most interesting regime is the inflationary
one. During inflation the system ``forgets" its initial conditions and
other characteristics of the pre-inflationary era, such as the possible
presence of other types of matter in addition to the scalar field,
spatial curvature, etc., due to a rapid growth of the scale factor. This
feature enables us to use such a simple model for describing the physics
of the early Universe from the instant when the inflationary regime was
established.

On the other hand, the problem of pre-in\-fla\-ti\-on\-ary era and initial
condition for inflation is for the same reason very difficult
because nowadays we have no physical ``probe" which might give us
information about that time. In such a situation we may hope to
extract some information from mathematical studies of the corresponding
dynamical system.  One of the most important problems is to describe the
set of initial conditions which led to the inflationary regime. It can
also clarify whether this regime is natural for this dynamical system
or it requires some kind of fine tuning of the initial data.

For this question to make sense, it is necessary to specify a measure on the
space of initial conditions. It is common to use the hypersurface with  the
energy density equal to the Planckean one (called the Planck boundary) as
the initial-condition space. A common angular measure on the Planck boundary
will be described below. As was pointed out in Ref.\,\cite{BKH} (see also a
detailed discussion of possible choices of the measure in the cited paper),
this choice is based on the physical considerations of inapplicability of
classical gravity beyond the Planck boundary and of the absence of
any information from this region. When the measure is specified, the
inflation generality problem can be studied quantitatively.

A solution can
depend on the physical condition in the epoch followed by inflation.
In Refs.\,\cite{four,Sato,B-Kh} it was found that the distribution
of initial data leading
to inflation strongly depends on the sign of the spatial curvature. If
it is negative or zero, the scale factor of the
Universe cannot pass through
extremum points (see below). In this case all the
trajectories in the configuration space ($a, \varphi$), where $a$ is the scale
factor and $\varphi$ is the scalar field, starting from
a sufficiently large value $\varphi_0$, reach a slow-roll regime and experience
inflation. If we start from the Planck energy, a measure of
non-inflating trajectories is about $m/m_{P}$ where $m$ is the mass of
the scalar field and $m_{P}$ is the Planck mass. From observational
reasons, this ratio is about $10^{-5}$ \cite{R-S-V,Haw}
so almost all trajectories lead to the inflationary regime. But
positive spatial curvature allows a trajectory to have a point of maximal
expansion which  results in increasing the measure of non-inflating
trajectories to $\sim 0.3$ \cite{Sato,B-Kh}.

Another important characteristic of the pre-in\-fla\-ti\-on\-ary era which
is also ``forgotten" during inflation is the possible presence of a
hydrodynamical matter in addition to the scalar field. Decreasing even more
rapidly than the curvature with increasing scale factor $a$, the energy
density of hydrodynamical matter could not affect the slow-rolling
conditions and so has a tiny effect on the dynamics if the spatial curvature
is nonpositive.  But the conditions for extrema of $a$ can change
significantly in a closed model, so the latter with a scalar field and
hydrodynamical matter requires special analysis.

The structure of this paper is as follows: in Sec. 2 we
consider the dynamical system corresponding to an isotropic Universe
with a massive scalar field and without any other type of matter.
We also show how to use the configuration space $(a, \varphi)$ for
illustrating the generality of inflation problem. In Sec. 3
this method is applied to a closed isotropic Universe with
scalar field and hydrodynamical matter in the form of perfect fluid. In
Sec. 4 we briefly discuss the generality of the results obtained in
Sec. 3.

\section {Basic equations and dimensionless variables}

We consider a cosmological model with the action
\begin{equation}
S = \int d^{4} x \sqrt{-g}\left\{\frac{m_{P} ^{2}}{16\pi} R +
\frac{1}{2} g^{\mu\nu}\partial_{\mu}\varphi \partial_{\nu}\varphi
-\frac{1}{2}m^{2}\varphi^{2}\right\}.
\end{equation}
For a closed Friedmann model with the metric
\begin{equation}
    ds^{2} =  dt^{2} - a^{2}(t) d^{2} \Omega^{(3)},
\end{equation}
where
$a(t)$ is the scale factor,
$d^{2} \Omega^{(3)}$ is the metric on a unit 3-sphere and $\varphi$
is a homogeneous scalar field, we can get the following equations of
motion
\begin{equation}
\frac{m_{P} ^{2}}{16 \pi}\left(\ddot{a} + \frac{\dot{a}^{2}}{2 a}
+ \frac{1}{2 a} \right)
+\frac{a \dot{\varphi}^{2}}{8} -\frac{m^{2} \varphi^{2} a}{8} = 0
\end{equation}
\begin{equation}
\ddot{\varphi} + \frac{3 \dot{\varphi}\dot{a}}{a} + m^2\varphi =
0.
 \end{equation}
 Besides, we can write down the first integral
of motion for our system
\begin{equation}
 -\frac{3}{8 \pi} m_{P} ^{2}
(\dot{a}^{2} + 1) +\frac{a^{2}}{2}\left(\dot{\varphi}^{2} + m^{2}
\varphi^{2}\right)=0.
 \end{equation}

It is easily seen from (2.5) that the points of maximal expansion
and contraction, i.e. the points where $\dot{a} =0$ can exist
only in a region where
\begin{equation}
  \varphi^{2} \leq \frac{3}{4 \pi} \frac{m_{P} ^{2}}{m^{2} a^{2}}, 
\end{equation}
which represents the field in the half-plane
$0 \leq a < +\infty$, $-\infty < \varphi <+\infty$
bounded by the hyperbolic curves
\[
\varphi \leq \sqrt{\frac{3}{4 \pi}} \frac{m_{P} }{m a}\quad {\bf and}\quad
\varphi \geq -\sqrt{\frac{3}{4 \pi}} \frac{m_{P} }{m a}
\]
(see Fig.\,1).
Sometimes the region determined by the inequalities (2.6) is called Euclidean
or ``classically forbidden''. One can argue about the validity of such a
definition (for details see \cite{our1}), but we shall use it for
convenience. Now we would like to distinguish between the maximal
contraction points where $\dot{a} = 0, \ddot{a} > 0$ and those of maximal
expansion where $\dot{a} = 0, \ddot{a} < 0$. Let us put
$\dot{a} = 0$, in this case one can express $\dot{\varphi}^{2}$ from
(2.5) as
\begin{equation}
    \dot{\varphi}^{2} = \frac{3}{4 \pi} \frac{m_{P} ^{2}}{a^{2}}
        -m^{2} \varphi^{2}.
\end{equation}
Substituting (2.7) and $\dot{a} = 0$ into Eq. (2.3), we have
\begin{equation}
    \ddot{a} = \frac{4 \pi m^{2} \varphi^{2} a}{m_{P} ^{2}}-\frac{2}{a}.
\end{equation}
From  (2.8) one can easily see that the possible points of
maximal expansion are localized inside the region
\begin{equation}
\varphi^{2} \leq \frac{1}{2 \pi} \frac{m_{P} ^{2}}{m^{2} a^{2}},
\end{equation}
while those of maximal contraction (bounces) lie outside
the region (2.9) being at the same time inside the Euclidean region
(2.6) (see Fig.\,1) \cite{our2}.

It is convenient to employ the dimensionless quantities
\[
   x=\frac{\varphi}{m_{P}},\qquad y=\frac{\dot \varphi}{m m_{P}},\qquad
z=\frac{\dot a}{m a}.  \]
 We will study the dynamics of the Universe
 starting from the Planck boundary
\begin{equation}
 \frac{m^2
\varphi^2}{2} + \frac{\dot \varphi^2}{2} = m_{P}^4
 \label{PB}
 \end{equation}
    We also introduce the convenient angular parametrization of the
Planck boundary
 \begin{equation}
\frac{m^2 \varphi^2}{2}=m_{P}^4
\cos^2{\phi}, \qquad \frac{\dot \varphi^2}{2}=m_{P}^4 \sin^2{\phi}.
\end{equation}

This variable $\phi$ along with the variable $z$ determine the initial
point on the Planck boundary completely. $z$ can vary in a compact region
from $0$ to $z_{\max}=\sqrt{8 \pi/3}(m_{P}/m)$;
the corresponding initial values of the scale factor $a$ vary from
$a_{\min}=\sqrt{3/(8 \pi m_{P}^2)}$ to $+\infty$.

The measure we have used is the area $S$ of the initial conditions on the
$(z, \phi)$ plane over the total area $S_{\rm Pl}$ of the Planck
boundary.

All definitions and properties of these variables remain unchanged
in the presence of ordinary matter [except for adding the matter energy in
the left-hand side of (\ref{PB})], which will be studied in the next
section.  And now we briefly recall the situation without any matter in
addition to the massive scalar field.  Though all plots will be in the
$(z,\phi)$-plane, it is useful to keep in mind the $(a,\varphi)$
configuration space.

As we know from \cite{our2}, the fate of a trajectory starting from $\dot
a=0$ can be deduced from the coordinates of the initial point: a trajectory
with $z=0$ and an initial point lying between the two hyperbolae of Fig.\,1
will expand, while an initial point lying below the separating curve leads
to contraction instead of the inflationary regime. It is clear from
(2.6), (2.9) that the scalar field value on the Euclidean
boundary is $\sqrt{2/3}$ times the value of the scalar field on the
separating curve for a fixed value of the scale factor. In the angular
parametrization (2.11), the value of $\varphi_0$ lying on the separating curve
and separating these two regimes corresponds to $\phi = \arccos(\sqrt{2/3}) \sim
0.61$. An initial, sufficiently large positive Hubble constant gives a
the universe a chance to pass a ``dangerous" region of possible
maximal expansion points (to the left from the separating curve) and to
reach the inflationry regime. The resulting distribution of initial points on
the plane $(z,\phi)$ which do not lead to inflation is as shown in
Fig.\,2(a).  A total measure of such  trajectories is about $30 \%$ of the
plane $(z, \phi)$.

\section{A model with scalar field and perfect fluid}

Now we add a perfect fluid with the equation of state $P=\gamma E$. The
parameter $\gamma$ can in principle vary in the range $-1\le \gamma \le 1$.
In this paper the case $-1/3 < \gamma \le 1$ will be
considered.  This case contains all known kinds of matter in the form
of a perfect fluid except the cosmological constant. Three cases of
particular physical interest  are $\gamma=0$ (dust), $\gamma=1/3$
(ultrarelativistic matter) and $\gamma=1$ (massless scalar field).

The equation of motion are now
\begin{equation}
\frac{m_{P}^{2}}{16 \pi}\left(\ddot{a} + \frac{\dot{a}^{2}}{2 a}
    + \frac{1}{2 a} \right) +\frac{a \dot{\varphi}^{2}}{8}
-\frac{m^{2} \varphi^{2} a}{8}
            - \frac{Q}{12 a^{p+1}}(1-p) = 0
\end{equation}
\begin{equation}
\ddot{\varphi}
    + \frac{3 \dot{\varphi} \dot{a}}{a}+ m^{2} \varphi = 0.
\end{equation}
with the constraint
\begin{equation}
-\frac{3}{8 \pi} m_{P}^2 (\dot{a}^{2} + 1)
+\frac{a^{2}}{2}\left(\dot{\varphi}^{2} + m^{2} \varphi^{2}\right) +
                \frac{Q}{a^{p}} = 0.
\end{equation}
Here $p=1+3 \gamma$, $Q$ is a constant from the equation of motion for matter
which can be integrated in the form
\begin{equation}
    E a^{p+2} = Q = const.
\end{equation}

Before presenting the results of a numerical integration of
(3.1)--(3.3) let us make some qualitative statements.

The Euclidean region is now bounded from large values
of the field $\varphi$ and small values of the scale factor $a$ (see Fig.\,1).
The upper point of the Euclidean boundary $\varphi=\varphi_{\max}$
corresponds to $a^p=4 \pi (p+2) Q/(3m_{P}^2) $ and the
fact that there is no bounce for bigger values of the scalar field is
related to a transition from chaotic to regular types of dynamics,
described by (3.1)--(3.3) as we have shown in \cite{ournew}.

But for our present purposes it is important that the Euclidean region
disappears at
\begin{equation}
    a^p=\frac{8 \pi}{3 m_{P}^2} Q.
\end{equation}
We also need the equation for $\ddot a$ at the points of bounce:
\begin{equation}
    \ddot a=-\frac{2}{a} + \frac{4 \pi}{m_{P}^2} m^2 a \varphi^2 +
        \frac{4 \pi}{3 m_{P}^2} \frac{Q}{a^{p+1}} (4-p).
\end{equation}
Thus for $p=4$ (a massless scalar field) the separating curve is the same as
it was for the case without matter (Fig\,1(b)), while for $p=1$ and $p=2$
we have an additional term (Fig\,1(c)). It is necessary to notice that the
part of the separating curve beyond the Euclidian region is related to the
so-called Euclidean counterparts of the equations of motion studied in
quantum cosmology (see \cite{our1}) and will not be considered here.

\begin{figure}
\epsfxsize=\hsize
\centerline{{\epsfbox{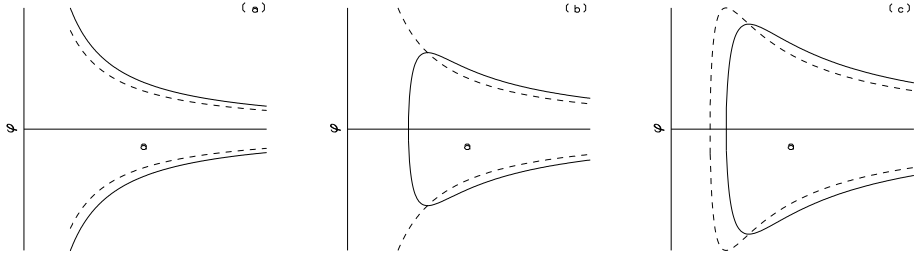}}}
\caption{
Configurations of the Euclidean boundary (solid) and the separating
curve (dashed), in Fig.\,1(a)  without hydrodynamical matter,
in Fig.\,1(b) in the presence of matter with $p=4$,
in Fig.\,1(c) in the presence of a fluid with $p \ne 4$.}
\end{figure}

In all cases the separating curve crosses the Euclidean boundary at
\begin{equation}
    a_{\rm cr}^p=\frac{4 \pi (2+p)}{3 m_{P}^2} Q,
\end{equation}
Keeping in mind this feature and the geometry of the Euclidean region, it is
sufficiently simple to explain the numerical results plotted in Fig.\,2.

The influence of matter is significant only for small values of the scale
factor which correspond to small values of $z$ in Fig.\,2. For $Q$ small
enough to keep $a_{\min}$ greater than $a_{\rm cr}$, the situation is like
that in Fig.\,2(b): only trajectories with small initial values of $z$
``feel" the presence of additional matter and can change their behaviour
significantly.  The measure of trajectories falling to singularity
slowly increases with increasing $Q$. When $a_{\min}=a_{\rm cr}$, all the
trajectories with zero velocity fall to the singularity because their
initial points lie lower than the separating curve (see Fig.\,2(c)).
A further increase of $Q$ leads to the situation of Fig.\,2(d) - there exists
some minimal value of $z$ which is necessary for reaching the inflationary
asymptotic. The measure of such trajectories keeps on diminishing with
increasing $Q$.

But for
\begin{equation}
    Q=(\frac{3}{8 \pi})^{1+p/2} m_{P}^{2-p}
\end{equation}
$a_{\min}$ becomes equal to the value (3.5). The Euclidean
boundary does not exist any more at $a_{\min}$. This situation corresponds
to a density of matter so large, that a large spatial curvature (small $z$)
is incompatible with the restriction of the initial density by the Planck
value.  The value $z_{\min}$ bounds the physically
admissible initial conditions with densities smaller than the Planck density
(Fig.\,2(e)). In such a situation the fraction of initial conditions
leading to inflation among all physically admissible initial conditions
($z>z_{\min}$) decreases with increasing $Q$ (see Figs.\,2(e) -- 2(f)).

\begin{figure}
\epsfysize=15cm
\centerline{{\epsfbox{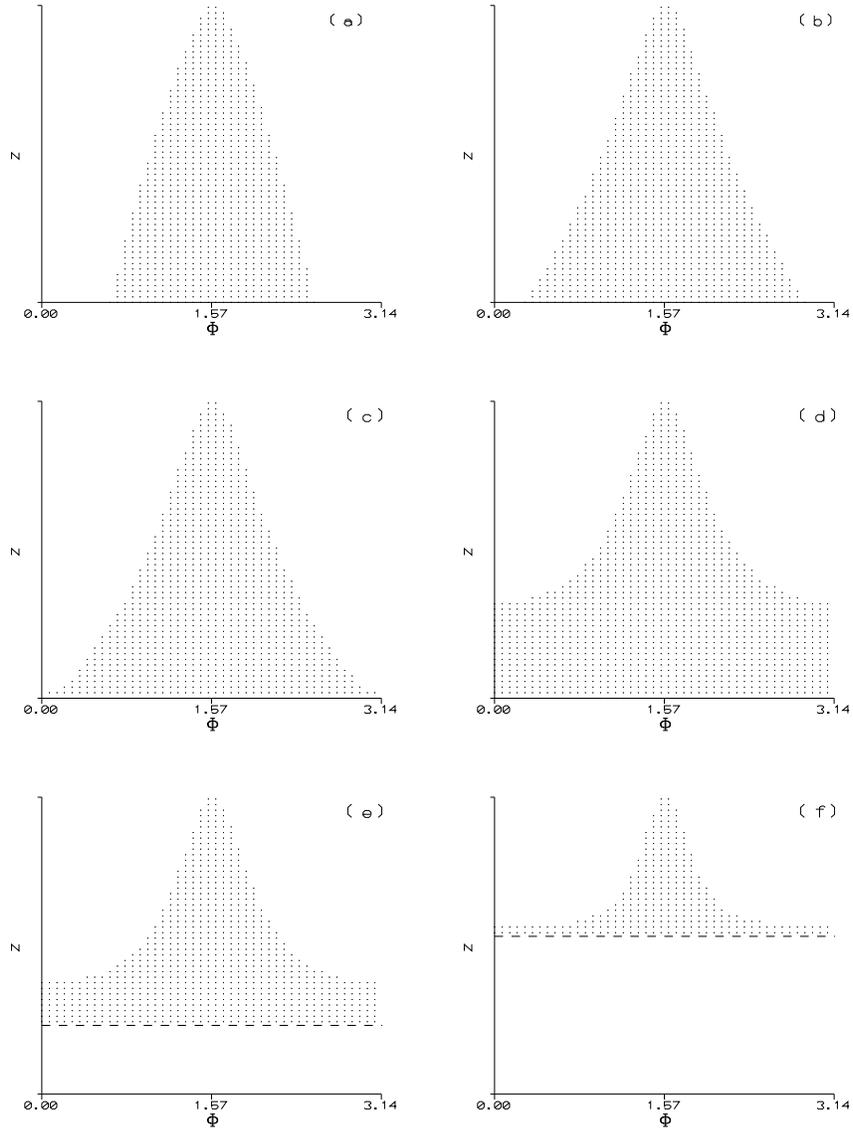}}}
\caption
{The area of initial data on the Planck boundary which do not lead
to inflation. Fig.\,2(a) corresponds to the situation without hydrodynamical
matter, other plots correspond to densities of the hydrodynamical matter
increasing  from Fig.\,2(b) to Fig.\,2(f) (see details in the text). The
value of $\phi$ varies is the range $ 0 \le \phi \le \pi$ and the value of
$z$ in the range $0 \le z \le z_{max}$.}
\end{figure}

So the measure of non-inflating trajectories as a function of the initial
density of ordinary matter has a maximum. This maximum corresponds
to such $Q$ that the density of ordinary matter at $z=0$ has just the
Planck value.  The numerical value of this maximum slowly depends on $p$.
For the three cases mentioned above the numerical values are

\begin{center}
$0.55$ for $p=1$ ($\gamma=0$),

$0.56$ for $p=2$ ($\gamma=1/3$),

$0.58$ for $p=4$ ($\gamma=1$).

\end{center}
\section{Discussion}

The results in \cite{Sato,B-Kh} are not essentially changed if we start not
exactly from the Planck energy but from smaller values. This is
important because we can not be sure that our model is valid
up to exactly the Planck energy scale. Indeed,
the essential criterion for inflation is that typical initial values of
$\varphi$ lie in the slow-rolling region. The value of $\varphi_{\rm sep}$
separating the slow-roll (for $\varphi>\varphi_{\rm sep}$)
 and oscillatory (for
$\varphi<\varphi_{\rm sep}$) regimes can be estimated as
$\varphi_{\rm sep}=m_{P}/(2 \sqrt{\pi})$.

Now, if we start from the energy density $E_{\rm in}=\epsilon m_{P}^4$,
$\epsilon<1$, the maximum possible initial value of $\varphi$ is
$\varphi=\sqrt{2 E_{\rm in}}/m$ and the condition $\varphi >> \varphi_{\rm sep}$
leads to $m^2/m_{P}^2 << \epsilon$. If this condition is satisfied,
than only a tiny part of the trajectories (with the measure
$\epsilon^{-1/2} m/m_{P}$) falls into an oscillatory regime with an
insuffitient degree of inflation while the main part of
non-inflationary trajectories falls into a singularity due to the
spatial curvature and their measure is almost independent of the scalar
field mass \cite{Sato}.

The presence of ordinary matter does not change the slow-roll regime, so
all the aforesaid about the influence of the scalar field mass on the
measure of non-inflationary trajectories is still valid.

The configuration of the Euclidean boundary and the separating curves in the
presence of matter depends on the value of the scale factor. Value of
the initial energy density deftermnes the minimal possible value of the scale
factor (see the constraint equation) and therefore can influence the results.
But it is clear from (3.5), (3.7) that the configuration of the curves
is invariant under transformations keeping ${Q}/{a^p}$ constant. Using
the equation for $a_{\min}$, this condition can be rewritten as
$Q\epsilon^{p/2}=const$.  Thus the transformations

$$
\begin{array}{l}
    E_{\rm in} \to \epsilon E_{\rm in}, \\
  Q \to \epsilon^{-p/2} Q
\end{array}
$$
leave the situation unchanged.  This symmetry of Euclidean and
separating curves indicates that the maximum fraction of non-inflating
trajectories does not depend on $\epsilon$. The maximum fraction is
achieved when the density of ordinary matter at $a_{\min}$ is equal to
$E_{\rm in}$.  These qualitative considerations were also confirmed by
direct numerical integration of the equations of motion.

As a result, the presence of a perfect fluid with $0\le\gamma\le 1$ in the
Universe filled by a massive scalar field can enlarge the fraction of
non-inflationary trajectories, but this fraction cannot exceed $\sim
60\%$ and the inflationary asymptotic remains rather natural.

\section*{Acknowledgments}
The author is grateful to A.Yu. Kamenshchik and I.M. Khalatnikov for
discussions and constant support.
This work was supported by Russian Basic Research Foundation via
grants 96-02-16220 and 96-02-17591.

\end {document}